\documentclass{article}
\usepackage{graphicx}
\usepackage{amsmath}

\newtheorem{theorem}{Theorem}

\newtheorem{case}[theorem]{Case}

\input{tcilatex}

\begin{document}

\title{Gravitational Frames and Scalar Field Dynamics}


\author{M. T. Ozaydin and N. Pirinccioglu}

\maketitle

\begin{abstract}
Scalar fields describe interesting phenomena such as Higgs bosons,
dark matter and dark energy, and are found to be quite common in
physical theories. These fields are susceptible to gravitational
forces so that being massless is not enough to remain conformal
invariant. They should also be connected directly to the scalar
curvature. Because of this characteristics, we investigated the
structure and interactions of scalar fields under the conformal
transformations. We show how to reduce the quadratic quantum
contributions in the single scalar field theory. In the multi-scalar
field theories, we analyzed interactions in certain limits. We
suggest a new method for stabilizing Higgs bosons.



\end{abstract}

\section{Introduction}\label{intro}

After the construction of general theory of relativity on Riemann
geometry by Einstein \cite{einstein1915}, the metric tensor which is
the same in everywhere of manifold became a basic geometric entitle.
Soon, with the analogy of gauge transformation of electromagnetism,
the idea of gauge transformation of geometry proposed by Weyl
\cite{weyl1918, weyl1922}. This idea was progressed by Cartan in
different studies \cite{cartan1922, cartan1923, cartan1924,
cartan1925}. Thus, the idea of two metric tensor of different
manifold connected via a conformal function was introduced by Weyl
\cite{weyl1922}. Then, the conformal transformation accumulated in
different studies in different context \cite{gursey1963, fubini1976,
englert1976, dicke1962, hoyle1974, demir2012}. For more information
one can see the \cite{carloni2009}, and references there in.
Recently, Demir \cite{demir2011} showed that the Higgs field could
be transferred from weak field to the gravitational field via a
conformal way. In general, scalar fields can be transferred in
metric sector by conformal way.

Experiments are still showing that the most working model for nature
is the Standard Model (SM) \cite{cern2013}. After the discovery of
Higgs boson, one missing part of the SM (electromagnetic, weak and
strong interactions) has been completed. Since the Higgs field is
scalar, naturally the quantum corrections of scalar fields are
unstable \cite{schwinger1970}. On the other hand, the experimental
evidences indicate the SM as more representing candidate of nature.
This indication leads us to the elimination instability of scalar
fields, which can be done using a conformal transformation. Nobili
\cite{nobili2012} show that the dilaton fields, which is geometry
based and vacuum expectation value of them are not zero, can
transfer energy from geometry to the matter sector by interacting
with the massless ordinary scaler fields, which has positive energy.
This interaction produces Higgs field. Thus, one important candidate
to combine the SM and general relativity (GR) is the conformal
general relativity (CGR).

In our study we show that unstable scalar field is to stabilized by
conformal way. This procedure is a new method for stabilizing Higgs
fields. In addition we analyzed two-scalar field in certain limits
using same method.

This paper is organized as follows. First section contains the
motivation and notation of the work. Second section contains the
Lagrangian with a scalar field and its conformal transformation.
Third section contains Lagrangian with two-scalar fields with
certain limits used. The conclusion is given in the last section.

\section{Notation and motivation}

After the idea of fields can attain mass via spontaneous symmetry
breaking mechanism, the CGR attract a new interest. Two metric
fields, $\widetilde{g}_{\mu\nu}$ and $g_{\mu\nu}$, which belong to
different manifolds can be related with smooth, strictly positive
function, \textit{i.e.} conformal function $\Omega(x)$ as follows:

\begin{equation}
\widetilde{g}_{\mu\nu}=\Omega^{2}g_{\mu\nu} \label{metric}.
\end{equation}

The inverse is

\begin{equation}
\widetilde{g}^{\mu\nu}=\Omega^{-2}g^{\mu\nu} \label{inversemetric},
\end{equation}

and its determinant is

\begin{equation}
\widetilde{g}=\Omega^{8}g \label{determinant}.
\end{equation}

The metric signature is $\hbox{diag}(-,+,+,+)$. In this regard, the
conformal transformation of  Cristoffel connection, Riemann tensor,
Ricci tensor, and Ricci scalar are respectively as follows
\cite{wald1984}.

\begin{eqnarray}
\widetilde{\Gamma}_{\mu\nu}^{\alpha}=
\Gamma_{\mu\nu}^{\alpha}+2\delta^{\alpha}_{(\mu}\nabla_{\nu)}\ln\Omega-g_{\mu\nu}g^{\alpha\beta}\nabla_{\beta}\ln\Omega
\label{cristoffel},
\end{eqnarray}

\begin{eqnarray}
\widetilde{R}_{\mu\nu\alpha}^{\lambda}=
R_{\mu\nu\alpha}^{\lambda}+2\delta^{\lambda}_{[\mu}\nabla_{\nu]}\nabla_{\alpha}\ln\Omega-2g^{\lambda\beta}g_{\alpha[\mu}\nabla_{\nu]}\nabla_{\beta}\ln\Omega+\nonumber\\
2(\nabla_{[\mu}\ln\Omega)\delta^{\lambda}_{\nu]}\nabla_{\alpha}\ln\Omega-2(\nabla_{[\mu}\ln\Omega)g_{\nu]\alpha}g^{\lambda\rho}\nabla_{\rho}\ln\Omega-\nonumber\\
2g_{\alpha[\mu}\delta^{\lambda}_{\nu]}g_{\beta\rho}\nabla_{\rho}\ln\Omega\label{riemann},
\end{eqnarray}

\begin{eqnarray}
\widetilde{R}_{\mu\alpha}=
R_{\mu\alpha}-(n-2)\nabla_{\mu}\nabla_{\alpha}\ln\Omega-g_{\mu\alpha}g^{\lambda\rho}\nabla_{\lambda}\nabla_{\rho}\ln\Omega+\nonumber\\
(n-2)(\nabla_{\mu}\ln\Omega)\nabla_{\alpha}\ln\Omega-(n-2)g_{\mu\alpha}g^{\lambda\rho}\nabla_{\lambda}\ln\Omega\nabla_{\rho}\ln\Omega
\label{ricci},
\end{eqnarray}

\begin{equation}
\widetilde{R}=
\Omega^{-2}\{R-2(n-1)g^{\mu\alpha}\nabla_{\mu}\nabla_{\alpha}
ln\Omega-(n-2)(n-1)g^{\mu\alpha}\nabla_{\mu}ln\Omega\nabla_{\alpha}ln\Omega\}
\label{curvature}.
\end{equation}

Where $\widetilde{R}=\widetilde{g}^{\mu\nu}\widetilde{R}_{\mu\nu}$.

\section{Conformal transformation in tensor-scalar fields}

An action of general theory with a scalar field can be written as
follows

\begin{equation}
S(\widetilde{g},\widetilde{\phi})=\int
d^{4}x\sqrt{-\widetilde{g}}\{\widetilde{C}(\widetilde{\phi})R(\widetilde{g})+
\widetilde{K}(\widetilde{\phi})\widetilde{g}^{\mu\nu}\partial_{\mu}\widetilde{\phi}
\partial_{\nu}\widetilde{\phi}-\widetilde{V}(\widetilde{\phi})\}.
\label{action1}
\end{equation}

Where $\widetilde{V}$ is potential and given by

\begin{equation}
\widetilde{V}(\widetilde{\phi})=\widetilde{V}_{0}+
\frac{1}{2}\widetilde{m}^{2}\widetilde{\phi}^{2}+
\frac{1}{4}\widetilde{\lambda}\widetilde{\phi}^{4},
\label{potential}
\end{equation}

$\widetilde{C}$ is conformal coupling given by

\begin{equation}
\widetilde{C}(\widetilde{\phi})=\frac{1}{2}(\widetilde{c}
\widetilde{\phi}^{2}+\widetilde{M}_{g}^{2}), \label{coupling}
\end{equation}

and $\widetilde{K}$ are functional coefficients,

\begin{equation}
\widetilde{K}(\widetilde{\phi})=\widetilde{k}
\end{equation}

$\widetilde{M}_{g}$ in equation (\ref{coupling}) is the
gravitational mass. Under the

\begin{equation}
\Omega(x)=\frac{M_{c}}{\widetilde{\phi}},
\end{equation}

\begin{equation}
\widetilde{g}^{\mu\nu}=\Omega^{-2}g^{\mu\nu},
\end{equation}

transformation, the action (\ref{action1}) becomes

\begin{eqnarray}
S(g,\phi)=\int d^{4}x\sqrt{-g}\{\frac{1}{2}(\widetilde{c}
M_{c}^{2}+(\frac{\widetilde{M}_{g}}{M_{c}})^{2}\phi^{2})R+
[3(\frac{\widetilde{M}_{g}}{M_{c}})^{2}+\nonumber\\
(\widetilde{k}-3\widetilde{c})
(\frac{M_{c}}{\phi})^{2}]g^{\mu\nu}\partial_{\mu}\phi\partial_{\nu}\phi-
\frac{1}{4}\widetilde{\lambda}M_{c}^{4}-\frac{1}{2}\widetilde{m}^{2}\phi^{2}-
\frac{\widetilde{V}_{0}}{M_{c}^{4}}\phi^{4}\}. \label{action2}
\end{eqnarray}

Where $M_{c}$ is the dimensional mass parameter. Setting
$\widetilde{c}M_{c}^{2}\equiv M_{pl}^{2}$,
$(\frac{\widetilde{M}_{g}}{M_{c}})^{2}\equiv -\xi_{c}$, and
$3\widetilde{c}=\widetilde{k}$, equation (\ref{action2}) becomes

\begin{equation}
S(g,\phi)=\int d^{4}x\sqrt{-g}
\{\frac{1}{2}[(M_{pl}^{2}-\xi_{c}\phi^{2})R-g^{\mu\nu}\partial_{\mu}\phi\partial_{\nu}\phi]-
V(\phi)\}. \label{action3}
\end{equation}

Where

\begin{equation}
V(\phi)=\frac{1}{4}\widetilde{\lambda}M_{c}^{4}+\frac{1}{2}\widetilde{m}^{2}\phi^{2}+
\frac{\widetilde{V}_{0}}{M_{c}^{4}}\phi^{4}, \label{potential2}
\end{equation}

$\xi_{c}=\frac{1}{6}=\widetilde{c}=\xi_{c},
\widetilde{M}_{g}^{2}=-M_{pl}^{2}$, and
$\phi=M_{c}^{2}/\widetilde{\phi}$. This conformal transformation has
two properties; firs, transformation function $\Omega$ is related to
the $\phi$. Second, scaler field $\phi$ transforms as
$\phi=M_{c}^{2}/\widetilde{\phi}$.

These two properties are emerging very important results:

1. From equation (\ref{potential}), the vacuum term
$\widetilde{V_{0}}$, transforms to the four-coupling parameter,
$\lambda$ in equation (\ref{potential2}) .

2. Four-coupling parameter $\widetilde{\lambda}$ in equation
(\ref{potential}), gives the vacuum energy $V_{0}$ in equation
(\ref{potential2}).

3. This conformal transformation does not change the scalar field
mass, $\widetilde{m}^{2}=m^{2}$ .

4. $\phi$ stays as a real scalar field, while $\widetilde{\phi}$ is
the ghost field.

5. From the transformation properties; even if equation
(\ref{action1}) has no scalar field, $\Omega$ treats as a new scalar
field and transforms the gravitational theory to a scalar tensor
theory.

These properties give a new way to overcame the problems of
stabilizing scalar field theories. The first one is the huge quantum
corrections to the scalar fields masses. In literature, \textit{
e.g.} in \cite{martin1997}, quantum correction of a scalar field
mass is given as

\begin{equation}
\delta\widetilde{m}^{2}\propto
\frac{\widetilde{\lambda}}{(4\pi)^{2}}\Lambda^{2}
\label{correction1}
\end{equation}

$\Lambda$ is the most reachable energy-momentum scale. In general it
is proportional to the $M_{pl}$. Quantum correction of mass is very
huge, and thus the theory become trivial at the quantum level. This
phenomena  works also for Higgs boson. But, under the conformal
transformation the quantum correction of (\ref{correction1}) can be
written as follows

\begin{equation}
\delta m^{2} \propto \frac{\lambda}{(4\pi)^{2}}\Lambda^{2}
\label{correction2}
\end{equation}

via a simple calculation. In this equation the $\lambda$ is equal to
the $\widetilde{V}_{0}/9M_{pl}^{4}$. This value is restricted with
the vacuum energy of equation (\ref{potential}),
$\widetilde{V}_{0}$. Thus, the value of $\delta m^{2}$ can be
reduced accordingly. In general, $\phi$ could be steady by setting
$\widetilde{V}_{0} \ll M_{pl}^{4}$. For example if we set
$\widetilde{V}_{0}\cong m^{2}M_{pl}^{4}/\Lambda^{2}$,  the mass of
$\phi$ will be stabilized. With this mechanism (conformal
transformation) the scalar field mass is being stabilized by
gravitational force. Because of Higgs field is scalar field, this
transformation also works for stabilizing Higgs boson mass.

\section{Multi-scalar fields}

The procedure which is performed in the last section, can be applied
to the multi scalar fields theories. For instance, a general theory
with two scalar fields can be written as follows.

\begin{eqnarray}
S(\widetilde{g},\widetilde{\phi})=\int
d^{4}x\sqrt{-\widetilde{g}}\{\frac{1}{2}[(\widetilde{M}^{2}_{g}+\widetilde{c}_{1}\widetilde{\phi}^{2}_{1}
+\widetilde{c}_{2}\widetilde{\phi}^{2}_{2})\widetilde{R}-
\widetilde{k}_{1}\widetilde{g}^{\mu\nu}\partial_{\mu}\widetilde{\phi}_{1}\partial_{\nu}\widetilde{\phi}_{1}-\nonumber
\\
\widetilde{k}_{2}\widetilde{g}^{\mu\nu}\partial_{\mu}\widetilde{\phi}_{2}\partial_{\nu}\widetilde{\phi}_{2}-
\widetilde{k}_{12}\widetilde{g}^{\mu\nu}\partial_{\mu}\widetilde{\phi}_{1}\partial_{\nu}\widetilde{\phi}_{2}]-
\widetilde{V}(\widetilde{\phi}_{1},\widetilde{\phi}_{2})\}.
\label{action4}
\end{eqnarray}

Where $\widetilde{k}_{1}, \widetilde{k}_{2}, \widetilde{k}_{12},
\widetilde{c}_{1}$, and $\widetilde{c}_{2}$ are coupling constants,
and $\widetilde{V}$ is the potential determined as
\begin{equation}
\widetilde{V}(\widetilde{\phi}_{1},\widetilde{\phi}_{2})=\widetilde{V}_{0}+
\frac{1}{2}\widetilde{m}_{1}^{2}\widetilde{\phi}_{1}^{2}+\frac{1}{2}\widetilde{m}_{2}^{2}\widetilde{\phi}_{2}^{2}+
\frac{1}{4}\widetilde{\lambda}_{1}\widetilde{\phi}_{1}^{4}+\frac{1}{4}\widetilde{\lambda}_{2}\widetilde{\phi}_{2}^{4}+
\frac{1}{4}\widetilde{\lambda}_{12}\widetilde{\phi}_{1}^{2}\widetilde{\phi}_{2}^{2}.
\label{potential3}
\end{equation}

 Under the

\begin{equation}
\widetilde{g}^{\mu\nu}=(\frac{\widetilde{\phi}_{1}}{M_{c_{1}}})^{\alpha_{1}}(\frac{\widetilde{\phi}_{2}}{M_{c_{2}}})^{\alpha_{2}}g^{\mu\nu},
\end{equation}
transformation and setting
\begin{equation}
\widetilde{\phi}_{1}=M_{c_{1}}^{2}/\phi_{1},
\end{equation}
\begin{equation}
\widetilde{\phi}_{2}=M_{c_{2}}^{2}/\phi_{2},
\end{equation}

the equation (\ref{action4}) becomes

\begin{eqnarray}
S(g,\phi)=\int d^{4}x\sqrt{-g}\{\frac{1}{2}[C(\phi_{1},\phi_{2})R+
K_{1}(\phi_{1},\phi_{2})g^{\mu\nu}\partial_{\mu}\phi_{1}\partial_{\nu}\phi_{1}+\nonumber
\\
K_{2}(\phi_{1},\phi_{2})g^{\mu\nu}\partial_{\mu}\phi_{2}\partial_{\nu}\phi_{2}+
K_{12}(\phi_{1},\phi_{2})g^{\mu\nu}\partial_{\mu}\phi_{1}\partial_{\nu}\phi_{2}]-V(\phi_{1},\phi_{2})\}.\label{action5}
\end{eqnarray}

Where $\alpha_{1}$, and $\alpha_{2}$ are arbitrary constants.
Functional coefficients $C$, $K_{1}$, $K_{2}$ and $K_{12}$, and
potential $V(\phi_{1},\phi_{2})$ in equation (\ref{action5}) are
defined as follows:

\begin{eqnarray}
C(\phi_{1},\phi_{2})=\widetilde{M}_{g}^{2}(\frac{\phi_{1}}{M_{c_{1}}})^{\alpha_{1}}(\frac{\phi_{2}}{M_{c_{2}}})^{\alpha_{2}}+
\widetilde{c}_{1}M_{c_{1}}^{2}(\frac{\phi_{1}}{M_{c_{1}}})^{\alpha_{1}-2}(\frac{\phi_{2}}{M_{c_{2}}})^{\alpha_{2}}+\nonumber
\\
\widetilde{c}_{2}M_{c_{2}}^{2}(\frac{\phi_{1}}{M_{c_{1}}})^{\alpha_{1}}(\frac{\phi_{2}}{M_{c_{2}}})^{\alpha_{2}-2}
\end{eqnarray}

\begin{eqnarray}
K_{1}(\phi_{1},\phi_{2})=\frac{3}{2}[\alpha_{1}^{2}(\frac{\widetilde{M}_{g}}{M_{c_{1}}})^{2}
(\frac{\phi_{1}}{M_{c_{1}}})^{\alpha_{1}-2}(\frac{\phi_{2}}{M_{c_{2}}})^{\alpha_{2}}+\nonumber\\
\widetilde{c}_{1}\alpha_{1}(\alpha_{1}+4)(\frac{\phi_{1}}{M_{c_{1}}})^{\alpha_{1}}(\frac{\phi_{2}}{M_{c_{2}}})^{\alpha_{2}}+\nonumber\\
\widetilde{c}_{2}\alpha_{1}^{2}(\frac{M_{c_{2}}}{M_{c_{1}}})^{2}(\frac{\phi_{1}}{M_{c_{1}}})^{\alpha_{1}-2}(\frac{\phi_{2}}{M_{c_{2}}})^{\alpha_{2}+2}]-
\widetilde{k}_{1}(\frac{\phi_{1}}{M_{c_{1}}})^{\alpha_{1}-4}(\frac{\phi_{2}}{M_{c_{2}}})^{\alpha_{2}},
\end{eqnarray}

\begin{eqnarray}
K_{2}(\phi_{1},\phi_{2})=\frac{3}{2}[\alpha_{2}^{2}(\frac{\widetilde{M}_{g}}{M_{c_{2}}})^{2}
(\frac{\phi_{1}}{M_{c_{1}}})^{\alpha_{1}}(\frac{\phi_{2}}{M_{c_{2}}})^{\alpha_{2}-2}+\nonumber\\
\widetilde{c}_{2}\alpha_{2}(\alpha_{2}+4)(\frac{\phi_{1}}{M_{c_{1}}})^{\alpha_{1}}(\frac{\phi_{2}}{M_{c_{2}}})^{\alpha_{2}}+\nonumber\\
\widetilde{c}_{1}\alpha_{2}^{2}(\frac{M_{c_{1}}}{M_{c_{2}}})^{2}(\frac{\phi_{1}}{M_{c_{1}}})^{\alpha_{1}+2}(\frac{\phi_{2}}{M_{c_{2}}})^{\alpha_{2}-2}]-
\widetilde{k}_{2}(\frac{\phi_{1}}{M_{c_{1}}})^{\alpha_{1}}(\frac{\phi_{2}}{M_{c_{2}}})^{\alpha_{2}-4},
\end{eqnarray}

\begin{eqnarray}
K_{12}(\phi_{1},\phi_{2})=
3[\frac{\alpha_{1}\alpha_{2}\widetilde{M}_{g}^{2}}{M_{c_{1}}M_{c_{2}}}(\frac{\phi_{1}}{M_{c_{1}}})^{\alpha_{1}-1}(\frac{\phi_{2}}{M_{c_{2}}})^{\alpha_{2}-1}+\nonumber\\
\widetilde{c}_{2}\alpha_{1}(\alpha_{2}+2)(\frac{M_{c_{2}}}{M_{c_{1}}})(\frac{\phi_{1}}{M_{c_{1}}})^{\alpha_{1}-1}(\frac{\phi_{2}}{M_{c_{2}}})^{\alpha_{2}+1}+\nonumber\\
\widetilde{c}_{1}\alpha_{2}(\alpha_{1}+2)(\frac{M_{c_{1}}}{M_{c_{2}}})(\frac{\phi_{1}}{M_{c_{1}}})^{\alpha_{1}+1}(\frac{\phi_{2}}{M_{c_{2}}})^{\alpha_{2}-1}]-
\widetilde{k}_{12}(\frac{\phi_{1}}{M_{c_{1}}})^{\alpha_{1}-2}(\frac{\phi_{2}}{M_{c_{2}}})^{\alpha_{2}-2},
\end{eqnarray}

and

\begin{eqnarray}
V(\phi_{1},\phi_{2})=\widetilde{V}_{0}(\frac{\phi_{1}}{M_{c_{1}}})^{2\alpha_{1}}(\frac{\phi_{2}}{M_{c_{2}}})^{2\alpha_{2}}+
\frac{1}{2}\widetilde{m}_{1}^{2}M_{c_{1}}^{2}(\frac{\phi_{1}}{M_{c_{1}}})^{2\alpha_{1}-2}(\frac{\phi_{2}}{M_{c_{2}}})^{2\alpha_{2}}+\nonumber\\
\frac{1}{2}\widetilde{m}_{2}^{2}M_{c_{2}}^{2}(\frac{\phi_{1}}{M_{c_{1}}})^{2\alpha_{1}}(\frac{\phi_{2}}{M_{c_{2}}})^{2\alpha_{2}-2}+
\frac{1}{4}\widetilde{\lambda}_{1}M_{c_{1}}^{4}(\frac{\phi_{1}}{M_{c_{1}}})^{2\alpha_{1}-4}(\frac{\phi_{2}}{M_{c_{2}}})^{2\alpha_{2}}+\nonumber\\
\frac{1}{4}\widetilde{\lambda}_{2}M_{c_{2}}^{4}(\frac{\phi_{1}}{M_{c_{1}}})^{2\alpha_{1}}(\frac{\phi_{2}}{M_{c_{2}}})^{2\alpha_{2}-4}+
\frac{1}{4}\widetilde{\lambda}_{12}M_{c_{1}}^{2}M_{c_{2}}^{2}(\frac{\phi_{1}}{M_{c_{1}}})^{2\alpha_{1}-2}(\frac{\phi_{2}}{M_{c_{2}}})^{2\alpha_{2}-2}.
\label{potential4}
\end{eqnarray}

The values of these functional coefficients and the potential are
very different for different values of $\alpha$. Some cases are
calculated as follows:

\begin{case}
$\alpha_{1}=1$, and $\alpha_{2}=0$:
\end{case}

\begin{equation}
C(\phi_{1},\phi_{2})=\widetilde{M}_{g}^{2}(\frac{\phi_{1}}{M_{c_{1}}})+
\widetilde{c}_{1}M_{c_{1}}^{2}(\frac{M_{c_{1}}}{\phi_{1}})(\frac{\phi_{2}}{M_{c_{2}}})+
\widetilde{c}_{2}M_{c_{2}}^{2}(\frac{\phi_{1}}{M_{c_{1}}})(\frac{M_{c_{2}}}{\phi_{2}})^{2}
\end{equation}

\begin{equation}
K_{1}(\phi_{1},\phi_{2})=\frac{3}{2M_{c_{1}}}[5\widetilde{c}_{1}\phi_{1}+
\widetilde{c}_{2}\frac{\phi_{2}^{2}}{\phi_{1}}+\frac{\widetilde{M}^{2}_{g}}{\phi_{1}}]-
\widetilde{k}_{1}(\frac{M_{c_{1}}}{\phi_{1}})^{3}
\end{equation}

\begin{equation}
K_{2}(\phi_{1},\phi_{2})=-\widetilde{k}_{2}\frac{\phi_{1}}{M_{c_{1}}}(\frac{M_{c_{2}}}{\phi_{2}})^{4}
\end{equation}

\begin{equation}
K_{12}(\phi_{1},\phi_{2})=6\widetilde{c}_{2}\frac{\phi_{2}}{M_{c_{1}}}-
\widetilde{k}_{12}(\frac{M_{c_{1}}}{\phi_{1}})(\frac{M_{c_{2}}}{\phi_{2}})^{2}
\end{equation}

\begin{eqnarray}
V(\phi_{1},\phi_{2})=V_{0}+ \frac{1}{2}m_{1}^{2}\phi_{1}^{2}+
\frac{1}{2}\widetilde{m}_{2}^{2}M_{c_{2}}^{2}(\frac{\phi_{1}}{M_{c_{1}}})^{2}(\frac{M_{c_{2}}}{\phi_{2}})^{2}+
\frac{1}{4}\widetilde{\lambda}_{1}M_{c_{1}}^{4}(\frac{M_{c_{1}}}{\phi_{1}})^{2}+\nonumber\\
\frac{1}{4}\widetilde{\lambda}_{2}M_{c_{2}}^{4}(\frac{\phi_{1}}{M_{c_{1}}})^{2}(\frac{M_{c_{2}}}{\phi_{2}})^{4}+
\frac{1}{4}\widetilde{\lambda}_{12}M_{c_{1}}^{2}M_{c_{2}}^{2}(\frac{M_{c_{2}}}{\phi_{2}})^{2}.
\end{eqnarray}

Where $V_{0}=\frac{1}{2}\widetilde{m}_{1}^{2}M_{c_{1}}^{2}$, the
transformed vacuum energy, and
$m_{1}^{2}=\frac{2\widetilde{V}_{0}}{M_{c_{1}}^{2}}$. Untransformed
vacuum energy $\widetilde{V}_{0}$ determines the transformed scalar
field $\phi_{1}$ mass, $m_{1}$.

\begin{case}
$\alpha_{1}=1$, and $\alpha_{2}=1$:
\end{case}

\begin{equation}
C(\phi_{1},\phi_{2})=\widetilde{M}_{g}^{2}(\frac{\phi_{1}}{M_{c_{1}}})(\frac{\phi_{2}}{M_{c_{2}}})+
\widetilde{c}_{1}M_{c_{1}}^{2}(\frac{M_{c_{1}}}{\phi_{1}})(\frac{\phi_{2}}{M_{c_{2}}})+
\widetilde{c}_{2}M_{c_{2}}^{2}(\frac{\phi_{1}}{M_{c_{1}}})(\frac{M_{c_{2}}}{\phi_{2}})
\end{equation}

\begin{equation}
K_{1}(\phi_{1},\phi_{2})=\frac{3}{2M_{c_{1}}M_{c_{2}}}[5\widetilde{c}_{1}\phi_{1}\phi_{2}+
\widetilde{c}_{2}\frac{\phi_{2}^{3}}{\phi_{1}}+\widetilde{M}_{g}^{2}\frac{\phi_{2}}{\phi_{1}}]-
\widetilde{k}_{1}\frac{\phi_{2}}{M_{c_{2}}}(\frac{M_{c_{1}}}{\phi_{1}})^{3}
\end{equation}

\begin{equation}
K_{2}(\phi_{1},\phi_{2})=\frac{3}{2M_{c_{1}}M_{c_{2}}}[5\widetilde{c}_{2}\phi_{1}\phi_{2}+
\widetilde{c}_{1}\frac{\phi_{1}^{3}}{\phi_{2}}+\widetilde{M}_{g}^{2}\frac{\phi_{1}}{\phi_{2}}]-
\widetilde{k}_{2}\frac{\phi_{1}}{M_{c_{1}}}(\frac{M_{c_{2}}}{\phi_{2}})^{3}
\end{equation}

\begin{equation}
K_{12}(\phi_{1},\phi_{2})=\frac{3}{M_{c_{1}}M_{c_{2}}}[\widetilde{M}_{g}^{2}+
3\widetilde{c}_{2}\phi_{2}^{2}+ 3\widetilde{c}_{1}\phi_{1}^{2}]-
\widetilde{k}_{12}\frac{M_{c_{1}}}{\phi_{1}}\frac{M_{c_{2}}}{\phi_{2}}
\end{equation}

\begin{eqnarray}
V(\phi_{1},\phi_{2})=V_{0}+
\frac{1}{2}m_{1}^{2}\phi_{1}^{2}+\frac{1}{2}m_{2}^{2}\phi_{2}^{2}+
\frac{1}{4}\widetilde{\lambda}_{1}M_{c_{1}}^{4}(\frac{M_{c_{1}}}{\phi_{1}})^{2}(\frac{\phi_{2}}{M_{c_{2}}})^{2}+\nonumber\\
\frac{1}{4}\widetilde{\lambda}_{2}M_{c_{2}}^{4}(\frac{\phi_{1}}{M_{c_{1}}})^{2}(\frac{M_{c_{2}}}{\phi_{2}})^{2}+
\frac{1}{4}\lambda_{12}\phi_{1}^{2}\phi_{2}^{2}.
\end{eqnarray}

Where
$\lambda_{12}=\frac{4\widetilde{V}_{0}}{(M_{c_{1}}M_{c_{2}})^{2}}$,
and
$V_{0}=\frac{1}{4}\widetilde{\lambda}_{12}(M_{c_{1}}M_{c_{2}})^{2}$.
The coupling constant, $\widetilde{\lambda}_{12}$, is transformed to
the vacuum energy, $V_{0}$, and the vacuum energy,
$\widetilde{V}_{0}$, is transformed to the coupling constant
$\lambda_{12}$. The scalar field masses are transformed as follows

\begin{equation}
m_{1}^{2}=\widetilde{m}_{2}^{2}(\frac{M_{c_{2}}}{M_{c_{1}}})^{2},
\end{equation}

\begin{equation}
m_{2}^{2}=\widetilde{m}_{1}^{2}(\frac{M_{c_{1}}}{M_{c_{2}}})^{2}.
\end{equation}

The scalar field masses are related to the $\widetilde{m}_{1},
\widetilde{m}_{2}, M_{c_{1}}$, and $M_{c_{2}}$.

\begin{case}
$\alpha_{1}=2$, and $\alpha_{2}=0$:
\end{case}

\begin{equation}
C(\phi_{1},\phi_{2})=\widetilde{M}_{g}^{2}(\frac{\phi_{1}}{M_{c_{1}}})^{2}+
\widetilde{c}_{1}M_{c_{1}}^{2}+\widetilde{c}_{2}M_{c_{2}}^{2}(\frac{\phi_{1}}{M_{c_{1}}})^{2}(\frac{M_{c_{2}}}{\phi_{2}})^{2}
\end{equation}

\begin{equation}
K_{1}(\phi_{1},\phi_{2})=\frac{6}{M_{c_{1}}^{2}}[3\widetilde{c}_{1}\phi_{1}^{2}+
\widetilde{M}_{g}^{2}+ \widetilde{c}_{2}\phi_{2}^{2}]-
\widetilde{k}_{1}(\frac{M_{c_{1}}}{\phi_{1}})^{2}
\end{equation}

\begin{equation}
K_{2}(\phi_{1},\phi_{2})=-\widetilde{k}_{2}(\frac{\phi_{1}}{M_{c_{1}}})^{2}(\frac{M_{c_{2}}}{\phi_{2}})^{4}
\end{equation}

\begin{equation}
K_{12}(\phi_{1},\phi_{2})=\frac{12}{M_{c_{1}}^{2}}\phi_{1}\phi_{2}-
\widetilde{k}_{12}(\frac{M_{c_{2}}}{\phi_{2}})^{2}
\end{equation}

\begin{eqnarray}
V(\phi_{1},\phi_{2})=V_{0}+ \frac{1}{2}m_{1}^{2}\phi_{1}^{2}+
\frac{1}{2}\widetilde{m}_{2}^{2}M_{c_{2}}^{2}(\frac{\phi_{1}}{M_{c_{1}}})^{4}(\frac{M_{c_{2}}}{\phi_{2}})^{2}+
\frac{1}{4}\lambda_{1}\phi_{1}^{4}+\nonumber\\
\frac{1}{4}\widetilde{\lambda}_{2}M_{c_{2}}^{4}(\frac{\phi_{1}}{M_{c_{1}}})^{4}(\frac{M_{c_{2}}}{\phi_{2}})^{4}+
\frac{1}{4}\widetilde{\lambda}_{12}M_{c_{2}}^{4}(\frac{\phi_{1}}{\phi_{2}})^{2}.
\end{eqnarray}

Where $\lambda_{1}=\frac{4\widetilde{V}_{0}}{M_{c_{1}}^{4}}$,
$V_{0}=\frac{1}{4}\widetilde{\lambda}_{1}M_{c_{1}}^{4}$, and
$\widetilde{m}_{1}=m_{1}$.

\begin{case}
$\alpha_{1}=2$, and $\alpha_{2}=2$:
\end{case}

\begin{equation}
C(\phi_{1},\phi_{2})=\widetilde{M}_{g}^{2}(\frac{\phi_{1}}{M_{c_{1}}})^{2}(\frac{\phi_{2}}{M_{c_{2}}})^{2}+
c_{1}\phi_{1}^{2}+ c_{2}\phi_{2}^{2}
\end{equation}

\begin{eqnarray}
K_{1}(\phi_{1},\phi_{2})=\frac{6}{M_{c_{1}}^{2}M_{c_{2}}^{2}}[\widetilde{M}_{g}^{2}\phi_{2}^{2}+
3\widetilde{c}_{1}\phi_{1}^{2}\phi_{2}^{2}+\widetilde{c}_{2}\phi_{2}^{4}]-
\widetilde{k}_{1}(\frac{M_{c_{1}}\phi_{2}}{M_{c_{2}}\phi_{1}})^{2}
\end{eqnarray}

\begin{eqnarray}
K_{2}(\phi_{1},\phi_{2})=\frac{6}{M_{c_{1}}^{2}M_{c_{2}}^{2}}[\widetilde{M}_{g}^{2}\phi_{1}^{2}+
3\widetilde{c}_{2}\phi_{1}^{2}\phi_{2}^{2}+\widetilde{c}_{1}\phi_{1}^{4}]-
\widetilde{k}_{2}(\frac{M_{c_{2}}\phi_{1}}{M_{c_{1}}\phi_{2}})^{2}
\end{eqnarray}

\begin{eqnarray}
K_{12}(\phi_{1},\phi_{2})=\frac{12}{M_{c_{1}}^{2}M_{c_{2}}^{2}}[\widetilde{M}_{g}^{2}\phi_{1}\phi_{2}+
2\widetilde{c}_{2}\phi_{1}\phi_{2}^{3}+2\widetilde{c}_{1}\phi_{2}\phi_{1}^{3}]-
\widetilde{k}_{12}
\end{eqnarray}

\begin{equation}
V(\phi_{1},\phi_{2})=V_{0}+
\frac{1}{2}m_{1}^{2}\phi_{1}^{2}+\frac{1}{2}m_{2}^{2}\phi_{2}^{2}+
\frac{1}{4}\lambda_{1}\phi_{1}^{4}+\frac{1}{4}\lambda_{2}\phi_{2}^{4}+
\frac{1}{4}\lambda_{12}\phi_{1}^{2}\phi_{2}^{2}.
\end{equation}

Where
$V_{0}=\widetilde{V}_{0}(\frac{\phi_{1}}{M_{c_{1}}})^{4}(\frac{\phi_{2}}{M_{c_{2}}})^{4}$,
$c_{2}=\widetilde{c}_{1}(\frac{M_{c_{1}}}{M_{c_{2}}})^{2}$,
$c_{1}=\widetilde{c}_{2}(\frac{M_{c_{2}}}{M_{c_{1}}})^{2}$,
$\lambda_{2}=\widetilde{\lambda}_{1}(\frac{M_{c_{1}}}{M_{c_{2}}})^{4}$,
$\lambda_{1}=\widetilde{\lambda}_{2}(\frac{M_{c_{2}}}{M_{c_{1}}})^{4}$,
$\widetilde{\lambda}_{12}=\lambda_{12}$.

The masses of scalar fields transform as

\begin{equation}
m_{1}^{2}=\widetilde{m}_{1}^{2}(\frac{\phi_{2}}{M_{c_{2}}})^{4},
\label{mass1}
\end{equation}

\begin{equation}
m_{2}^{2}=\widetilde{m}_{2}^{2}(\frac{\phi_{1}}{M_{c_{1}}})^{4}.\label{mass2}
\end{equation}

These are the some cases of $\alpha$'s. One can choice the suitable
cases for the relevant physical problems. Not only for stabilizing
the Higgs boson, it may useful in representing the dark energy as
well.

In the all cases;

1. Even if the action has no scalar field, $\Omega$ treats as new
scalar fields and transform the gravitational theory to a scalar
tensor theory.

2. All kinetic terms in the theory couple to the fields, $\phi_{1}$
and $\phi_{2}$, via functional couplings $K_{1}(\phi_{1},\phi_{2}),
K_{2}(\phi_{1},\phi_{2})$, and $K_{12}(\phi_{1},\phi_{2})$.

3. In the case 4, the scalar field masses, and vacuum energy become
the function of scalar fields,  $\phi_{1}, \phi_{2}$, eqs.
(\ref{mass1}, \ref{mass2}).

\section{Conclusion}

Conformal transformation transforms ghosty scalar field to the real
scalar field in the single scalar field theories. This
transformation also helps to reduces the quadratic quantum
correction to the scalar field mass, and gives an important way to
make  Higgs boson, founded in LHC experiments, as a stable particle.

In the multi-scalar field section, the situations contain many
specific cases. These situations are directly related to the
conformal functional coefficients $K_{1}, K_{2},$ and $K_{12}$.
Under the suitable choice of these coefficients one may get the
desired results. These coefficients can be organized to modeling the
inflation, dark energy \emph{etc} as well. The coefficients can be
set to the negative values in same cases; for example case 1, and
case 2. This is directly related to the characteristic of the real
scaler fields. On the other hand, the positive value of the
coefficients is related to the geometrical character of the ghosty
scalar fields.

For all the values of $\alpha$'s which we considered; the vacuum
energy, and the masses of scalar fields are transform very
differently. In the last case, the masses of scalar fields and
vacuum energy are related to the scalar fields, $\phi_{1}$ and
$\phi_{2}$.

\section*{Acknowledgments}

The authors would like to thank D. A. Demir for inspiration of
research and useful discussion.

\end{document}